\documentclass[sn-mathphys,iicol]{sn-jnl}

\usepackage{graphicx}
\usepackage{wrapfig, blindtext}
\usepackage{booktabs}
\usepackage{siunitx}
\usepackage{lineno}

\jyear{2021}%

\theoremstyle{thmstyleone}%
\theoremstyle{thmstyletwo}%

\theoremstyle{thmstylethree}%

\begin{document}

\title[Anomaly Detection]{Comparison of Supervised and Unsupervised Anomaly Detection in Belle\,II Pixel Detector Data}


\author[1]{\fnm{Johannes} \sur{Bilk}}\email{Johannes.Bilk@physik.uni-giessen.de}
\author*[1]{\fnm{Katharina} \sur{Dort}}\email{Katharina.Dort@physik.uni-giessen.de}
\author[1]{\fnm{Stephanie} \sur{K\"as}}\email{Stephanie.Kaes@physik.uni-giessen.de}
\author[1]{\fnm{Jens S\"oren} \sur{Lange}}\email{Soeren.Lange@exp2.physik.uni-giessen.de}
\author[1]{\fnm{Marvin} \sur{Peter}}\email{Marvin.Peter@physik.uni-giessen.de}
\author[1]{\fnm{Timo} \sur{Schellhaas}}\email{Timo.Schellhaas@physik.uni-giessen.de}
\author[2]{\fnm{Benjamin} \sur{Schwenker}}\email{Benjamin.Schwenker@phys.uni-goettingen.de}
\author[3]{\fnm{Bj\"orn} \sur{Spruck}}\email{bspruck@uni-mainz.de}
 
\affil*[1]{\orgdiv{II. Physikalisches Institut}, \orgname{Justus-Liebig-University Gie{\ss}en}, \orgaddress{\street{Heinrich-Buff-Ring}, \city{Gie{\ss}en}, \postcode{35392}, \country{Germany}}}

\affil[2]{\orgdiv{II. Physikalisches Institut}, \orgname{Georg-August-Universität G\"ottingen}, \orgaddress{\street{Friedrich-Hund-Platz}, \city{G\"ottingen}, \postcode{37077}, \country{Germany}}}

\affil[3]{\orgdiv{Institut f\"ur Kernphysik}, \orgname{Johannes Gutenberg-Universit\"at Mainz}, \orgaddress{\street{Johann-Joachim-Becher-Weg}, \city{Mainz}, \postcode{55128}, \country{Germany}}}


\abstract{
Machine learning has become a popular instrument for the identification of dark matter candidates at particle collider experiments.
They enable the processing of large datasets and are therefore suitable to operate directly on raw data coming from the detector, instead of reconstructed objects.
Here, we investigate patterns of raw pixel hits recorded by the Belle II pixel detector, that is operational since 2019 and presently features 4\,M pixels and trigger rates up to 5\,kHz.
In particular, we focus on unsupervised techniques that operate without the need for a theoretical model. 
These model-agnostic approaches allow for an unbiased exploration of data, while filtering out anomalous detector signatures that could hint at new physics scenarios. 
We present the identification of hypothetical magnetic monopoles against Belle II beam background using Self-Organizing Kohonen Maps and Autoencoders.
The two unsupervised algorithms are compared to a convolutional Multilayer Perceptron and a superior signal efficiency is found at high background rejection levels. 
Our results strengthen the case for using unsupervised machine learning techniques to complement traditional search strategies at particle colliders and pave the way to potential online applications of the algorithms in the near future. 

}




\maketitle

\section{Introduction}
\label{intro}

Machine learning has proven to be a valuable tool in reconstruction and analysis tasks for High-Energy Physics (HEP)~\cite{albertsson2018machine}.
In particular, the classification of signal and background or \textit{particle identification} (PID) using machine learning algorithms has sparked significant interest in recent times. 
The majority of these algorithms are trained in a supervised manner, and therefore rely on a prior definition of signal provided by a theoretical framework and simulations.
However, detector signatures corresponding to elusive beyond the Standard Model Physics (BSM) processes might be missed owing to a narrow signal definition or a mis-modelling of either signal or background. 

Unsupervised and semi-supervised methods aim at the identification of signal features while minimizing predictions about signal or background.  
A data-driven approach is adopted allowing for a model-agnostic analysis that has the advantage of being independent from theoretical assumptions and therefore not confined to specific signal hypotheses and background modelling. 
In this document, we present two unsupervised machine learning methods with the objective to identify anomalous detector patterns that are potential indicators for unaccounted physics scenarios. 
The anomaly detection is performed such that it can be considered as a sophisticated filter mechanism that defines a potential signal region for further statistical analysis.
In this paper, we focus on the filter itself.

The inner regions of the Belle II detector is considered to demonstrate this filtering approach.
The input data consists of pixel hits coming from the Belle\,II pixel detector, presently featuring one layer of DEPFET silicon sensors.
The unsupervised, data-driven machine learning algorithms are trained on background data recorded by the pixel detector. 
Simulated pixel hits by hypothetical long-lived magnetic monopoles serve as anomalous events to evaluate the performance of the presented algorithms. 
The unsupervised techniques are compared to a convolutional neural network, that is trained in a supervised manner.

The paper is structured as follows: In Section~\ref{sec:belleII}, the Belle\,II experiment and the pixel detector are introduced. 
Subsequently, the dataset and data preprocessing are described in Section~\ref{sec:data}
The different machine learning algorithms and their performance are presented in Section~\ref{sec:ML}.

\begin{figure*}[tbp]
	\centering
	\includegraphics[width=\textwidth]{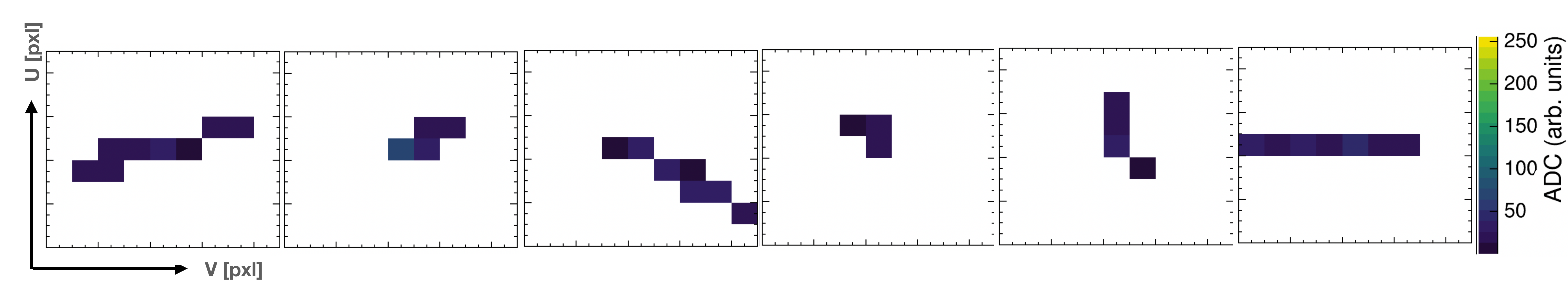}%
	\caption{Example clusters of single-beam background particles from the inner region of the PXD (pixel dimensions: \SI{50x55}{\micro \meter}).
		The seed pixel is located at the centre of the 9\,$\times$\,9 pixel matrix. 	
		The colour scale represents the single-pixel charge in ADC values. }
	\label{fig:clusters_background}
\end{figure*}
\begin{figure*}[tbp]
	\centering
	\includegraphics[width=\textwidth]{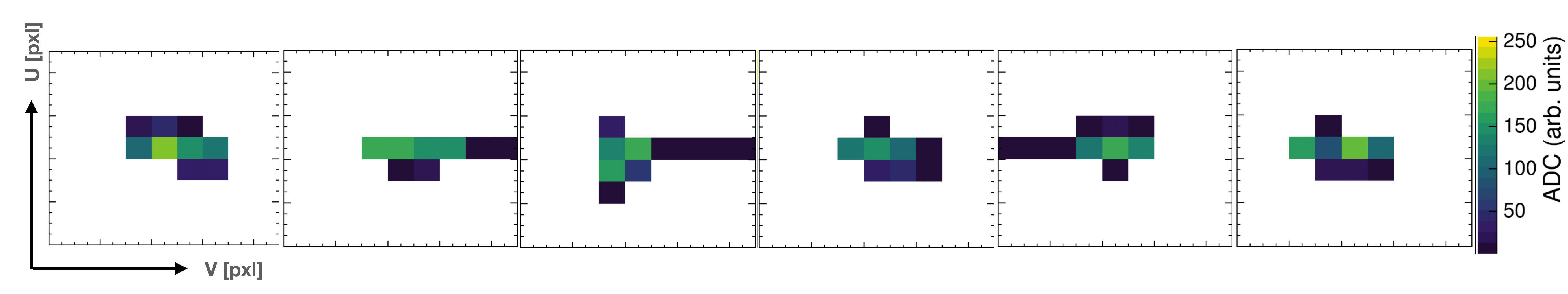}%
	\caption{Example clusters of simulated magnetic monopoles from the inner region of the PXD (pixel dimensions: \SI{50x55}{\micro \meter}).
		The seed pixel is located at the centre of the 9\,$\times$\,9 pixel matrix. 	
		The colour scale represents the single-pixel charge in ADC values. }
	\label{fig:clusters_monopoles}
\end{figure*}

\section{The Belle Experiment}
\label{sec:belleII}
The Belle II experiment, located at the SuperKEKB accelerator, has started operation in spring 2019. 
SuperKEKB provides electron-positron collisions with a nominal centre-of-mass energy of 10.58\,GeV and a design instant luminosity of $>\SI{5d35}{cm^2s^1}$~\cite{akai2018superkekb}.

The Belle II detector is composed of several sub-detectors, that are  arranged cylindrically around the interaction point~\cite{abe2010belle}. 
This document only treats the inner-most detector: an all-silicon pixel detector that is part of the Belle II tracking system, as detailed below.

\subsection{The Belle II Pixel Detector}
\label{sec:belleIIVertex}

The Belle II pixel detector (PXD) consists of pixelated DEPFET sensors, that are arranged in two layers at radii of 14\,mm and 22\,mm from the beam pipe~\cite{marinas2013belle,alonso2013depfet}. 
Presently, only the first layer is installed. 
The PXD features nearly 4 million pixels with pixel sizes between $\SI{50x55}{\micro m}^2$ and $\SI{50x85}{\micro m}^2$ and a thickness down to $\SI{75}{\micro m}$.

The data rate coming from the PXD is foreseen to reach about 20\,GB/s, which necessitates an online reduction scheme~\cite{levit2015fpga}. 
The FPGA-based data reduction system Online Selection Node (ONSEN) is able to reduce the data rate by a factor of 30 by using extrapolated reconstructed tracks, provided by the online event reconstruction, to the pixel detector layers and defining a region-of-interest (ROI) around the intercept of the track with the detector layers~\cite{gessler2015onsen}. 
However, this filtering mechanism relies on reconstructable particle tracks, as PXD hits outside a ROI are discarded by the ONSEN. 
In particular, particles with a low transverse momentum can escape tracking.
As a consequence, no ROI is generated and the PXD data associated with these particles is lost.

To guarantee high signal efficiency of particles with non-reconstructable tracks, a new veto system based on machine learning is proposed.
A proof-of-principal for a veto system dedicated to the identification of slow pions has already been presented in the past~\cite{bahr2015online}. 
In this document, the cluster rescue veto system is extended to exotic or \textit{anomalous} particle signatures that do not generate a reconstructable particle track. 
To assess the efficiency of the cluster rescue mechanism, we simulate the creation of long-lived hypothetical magnetic monopoles in the particle collision.
As a consequence of their high energy loss, magnetic monopoles are stopped in the inner layers of the Belle II detector~\cite{cecchini2016}.
The lack of hits in the outer sub-detectors inhibits the reconstruction of tracks, which also leads to the deletion of PXD data associated with a monopole by the ONSEN, once the ROI selection is switched on.
The aim of the proposed veto system is to identify the relevant PXD data based on anomalous event signatures and tagging it to prevent deletion. 
We consider unsupervised machine learning algorithms to generate the veto, that could potentially run online during data-taking on FPGA-based systems. 

\section{Data Generation}
\label{sec:data}

PXD background data was recorded in dedicated beam background runs taken in 2020.
For these runs, only a single particle beam circulated in the Belle II detector.
Background generation mechanisms such as the interaction of the circulating beam with residue particles in the beam pipe are responsible for the background hits detected in these runs. 
The PXD hits generated by background are characterised by a small charge signal in each pixel, as shown by the example clusters displayed in Fig~\ref{fig:clusters_background}.
The hits are nevertheless detected by the PXD due to the high signal-to-noise ratio of the DEPFET pixel sensors.  
The $v$ coordinate is along the beam direction and the $u$ coordinate perpendicular to it.
In view of future online applications of the investigated algorithms, the raw PXD hit information is used without applying an offline calibration. 

The signal events are simulated using the official Belle II framework \texttt{basf2}~\cite{kuhr2019belle}. 
The creation of monopole-antimonopole pairs from electron-positron collisions is considered.
The magnetic charge of the monopole is set to 68.5\,e in accordance with the Dirac theory~\cite{dirac1931quantised,milton2006theoretical} and the mass to 3\,GeV.
A full detector simulation with \texttt{basf2} is performed, including the interaction of particles with the PXD layers.
The simulated PXD hits associated with magnetic monopoles are shown in Fig.~~\ref{fig:clusters_monopoles}.

The PXD single-pixel hits that are used for the anomaly detection are extracted as follows:
For each simulated event and for the background events, the charge values of a 9\,$\times$\,9 pixel matrix are considered around the PXD hit with the highest charge value (\textit{seed pixel}). 
The matrix size is sufficiently large to capture the entire cluster extent for the majority of events and small enough to guarantee a fast convergence of the investigated algorithms.
In addition, the global position of the seed pixel is extracted. 
In total, 84 features are considered, which are normalized to the range [0, 1] to avoid dominance of a single parameter. 
Dimensionality reduction using a Principle Component Analysis has exhibited adverse  effects on the performance of the considered algorithms and is therefore not employed. 

\section{Machine learning techniques}
\label{sec:ML}

We propose a sophisticated filter based on unsupervised machine learning algorithms to identify anomalous signatures in the PXD. 
The filter operates on an event-by-event basis and labels each 9\,$\times$\,9 matrix from an event as anomalous or normal based on an \textit{anomaly score}.
While the scope of the anomaly score depends on the selected algorithm, we adopt the definition that low values represent normal and high values anomalous events.

\subsection{Performance metrics}

For all algorithms, the Receiver Operating Characteristics (ROC) is obtained by scanning the signal efficiency $\epsilon_S$ and recording the background rejection $\epsilon_B$. 
The Area-Under-Curve (AUC) is commonly used as a figure of merit for the performance of classifiers~\cite{hanley1982meaning}. 
For anomaly detection, a high background rejection is particularly desirable.
Therefore, the signal efficiency at three different operation points featuring a high background rejection level are studied as well, i.e. the signal efficiencies $\epsilon_S(\epsilon_B = 10^{-2})$, $\epsilon_S(\epsilon_B = 10^{-3})$ and $\epsilon_S(\epsilon_B = 10^{-4})$ are extracted. 

The uncertainty is extracted by repeating the training and evaluation five times with random shuffling of input vectors i.e. a vector in the training set in the first iteration can be assigned to the evaluation set in the second one. 
In each iteration, the performance metrics are determined.
Their mean represents the nominal value and the quadratic sum of deviations the uncertainty.  

\begin{figure}[tbp]
	\centering
	\includegraphics[width=\columnwidth]{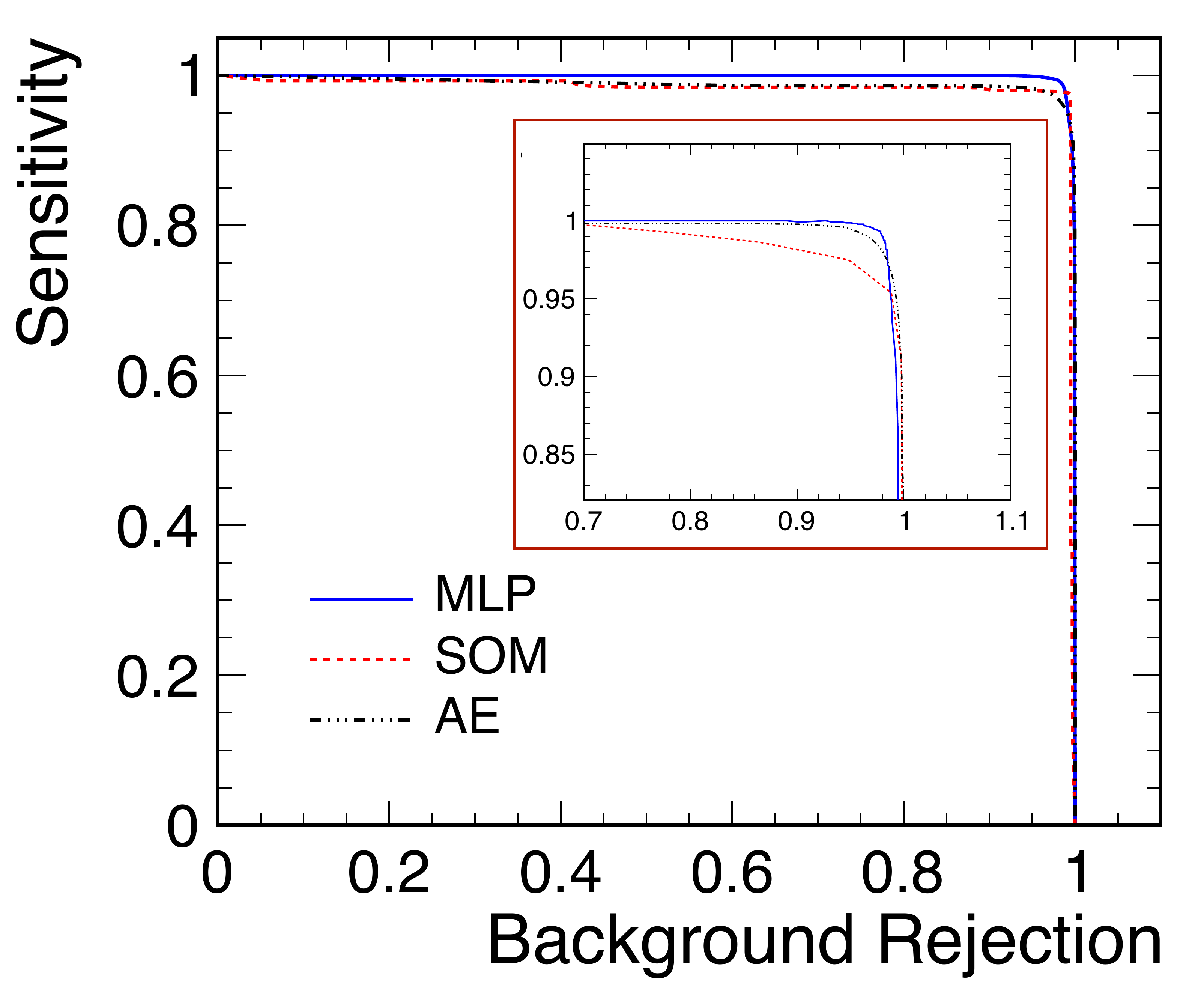}%
	\caption{ROC curves for different machine learning algorithms.}
	\label{fig:rocOverlay_202109}
\end{figure}


\begin{table*}[t]
	\centering
	\begin{tabular}{llllll}
		\toprule
		\textbf{Algorithm} & AUC [\%] & $\epsilon_S(10^{-1})$ [\%]  & $\epsilon_S( 10^{-2})$  [\%] & $\epsilon_S(10^{-3})$ [\%] & $\epsilon_S(10^{-4})$  [\%] \\ 
		\midrule
		MLP & $99.86^{+0.01}_{-0.02}$ & 100  & $97.3^{+0.2}_{-0.2}$  &  $82.4^{+1.2}_{-1.6}$ & $30.0^{+4.3}_{-3.8}$   \\
		SOM &  $98.15^{+0.04}_{-0.04}$ & $97.3^{+0.2}_{-0.2}$ & $94.0^{+0.8}_{-1.6}$  & $88.7^{+3.8}_{-5.4}$ &  $56.8^{+7.1}_{-8.8}$   \\
		AE & $98.86^{+0.05}_{-0.02}$ & $98.2^{+0.1}_{-0.3}$  &   $95.9^{+1.2}_{-0.4}$&  $87.4^{+1.4}_{-1.2}$ & $60.1^{+3.3}_{-2.7}$  \\
		\bottomrule
	\end{tabular}
	\caption{Performance metrics of the investigated machine learning techniques for the evaluation set.}
	\label{tab:performance}
\end{table*}

\subsection{Multilayer Perceptron (MLP)}

\begin{table*}[t]
	\centering
	\begin{tabular}{ll}
		\toprule
		\textbf{Parameter} & \textbf{Value}  \\ 
		\midrule
		Activation function & Rectified Linear Unit (ReLu) \\
		Loss function & Cross Entropy \\
		Optimizer & Stochastic Gradient Descent (SGD) \\
		Learning Rate & 1e-4 \\
		Momentum & 0.9 \\
		\bottomrule
	\end{tabular}
	\caption{Hyperparameters used for the Multilayer Perceptron.}
	\label{tab:parametersMLP}
\end{table*}

First, a supervised machine learning technique is considered, to which the unsupervised learning approaches are compared.
A convolutional Multilayer Perceptron (MLP) is utilized.
Details about the network architecture can be found in the Appendix.
The hyperparamters are listed in Table~\ref{tab:parametersMLP}.
The network is optimised for a high background rejection. 


The supervised training is performed using 350k background and signal events each.
After each training epoch, a dedicated testing set is presented to the MLP containing additional 150\,k events for both classes.
The training is stopped automatically once the reduction of the predicted error (\textit{loss}) from the testing set is only marginal. 

An evaluation set comprising 500\,k events for each class is considered to assess the performance of the algorithm.
The one-dimensional classification distribution for both classes is presented in Fig.~\ref{fig:rocOverlay_202109}. 
For low/high classification values the signal/background is suppressed by approximately three orders of magnitude. 
The corresponding ROC curve is shown in Fig.~\ref{fig:rocOverlay_202109} and the performance results are listed in Table~\ref{tab:performance}.
The AUC evaluates to $99.86^{+0.01}_{-0.02}$\,\%, which indicates an overall good classification performance. 
At high background rejection levels of $\epsilon_S(\epsilon_B = 10^{-4})$, the signal efficiency deteriorates to $30.0^{+4.3}_{-3.8}$\,\%. 

\begin{figure}[tbp]
	\centering
	\includegraphics[width=\columnwidth]{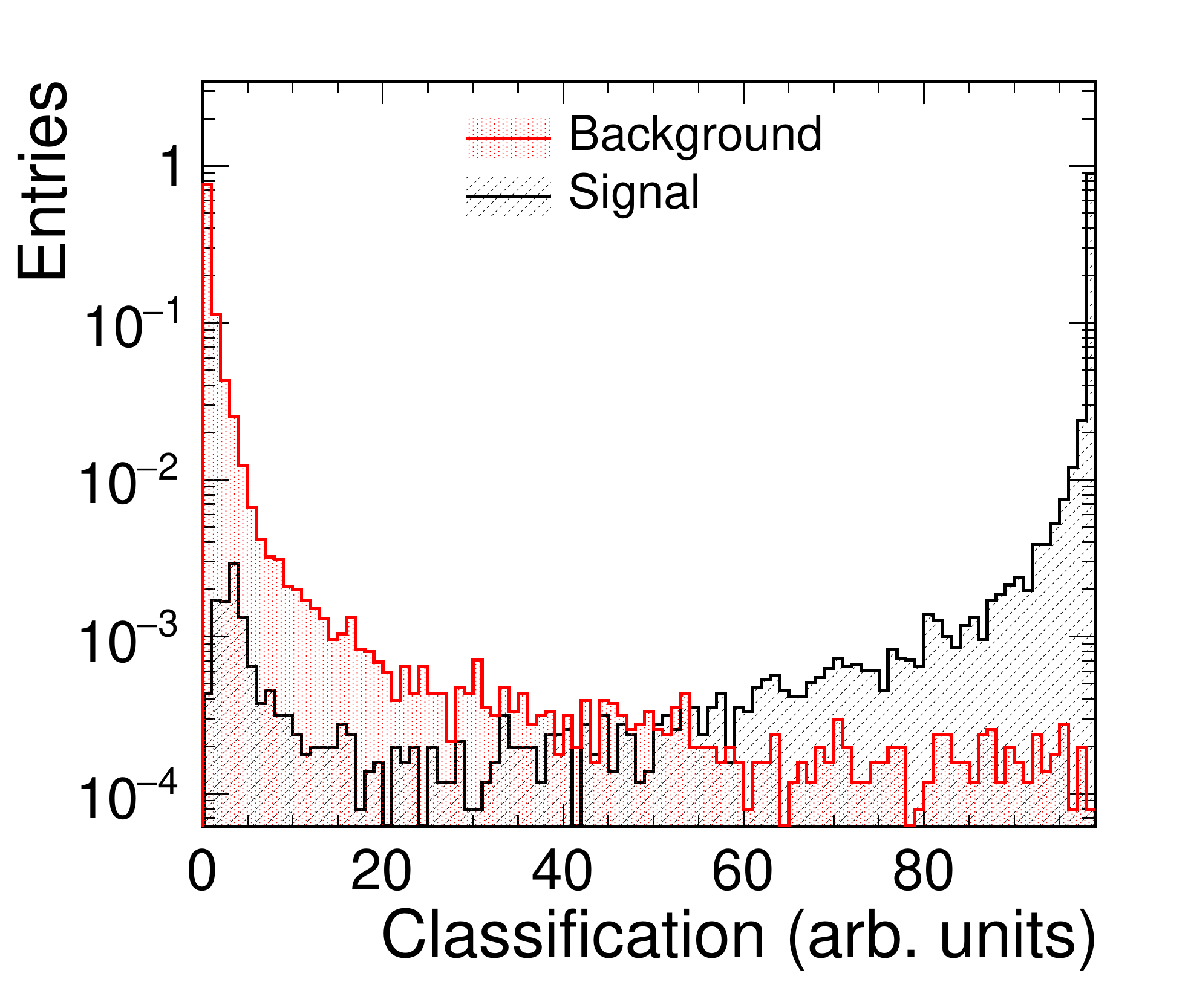}%
	\caption{MLP classification distribution for signal and background.}
	\label{fig:classification_mlp_202110}
\end{figure}


%
%
%

\subsection{Self-Organizing Maps (SOM)}

\begin{table*}[t]
	\centering
	\begin{tabular}{ll}
		\toprule
		\textbf{Parameter} & \textbf{Value}  \\ 
		\midrule
		Nodes & 70\\
		Neighbourhood function & Gaussian \\
		Gaussian width & 7 \\
		Learning Rate & 0.01 \\
		\bottomrule
	\end{tabular}
	\caption{Hyperparameters used for the Self-Organizing Maps}
	\label{tab:parametersSOM}
\end{table*}

A Self-Organizing Map (SOM) is an unsupervised machine learning technique, enabling the transformation of a high-dimensional dataset to a low-dimensional, discrete grid, while keeping the topological structure~\cite{kohonen1982self,kiviluoto1996topology}. 
After training, vectors that are close in the high-dimensional input space are represented by adjacent grid points in the low-dimensional space. 

The hyperparameters of the SOM are listed in Tab~\ref{tab:parametersSOM}.
The dimension of the low-dimensional grid space is set to one to allow for a comparable performance evaluation as for the other two machine learning techniques.
The low-dimensional representation will therefore span only a single line, with the aim to have grid points responding to background cluster on the one end and signal on the other end. 
The same training and testing sets as for the MLP are used. 

\begin{figure}[tbp]
	\centering
	\includegraphics[width=\columnwidth]{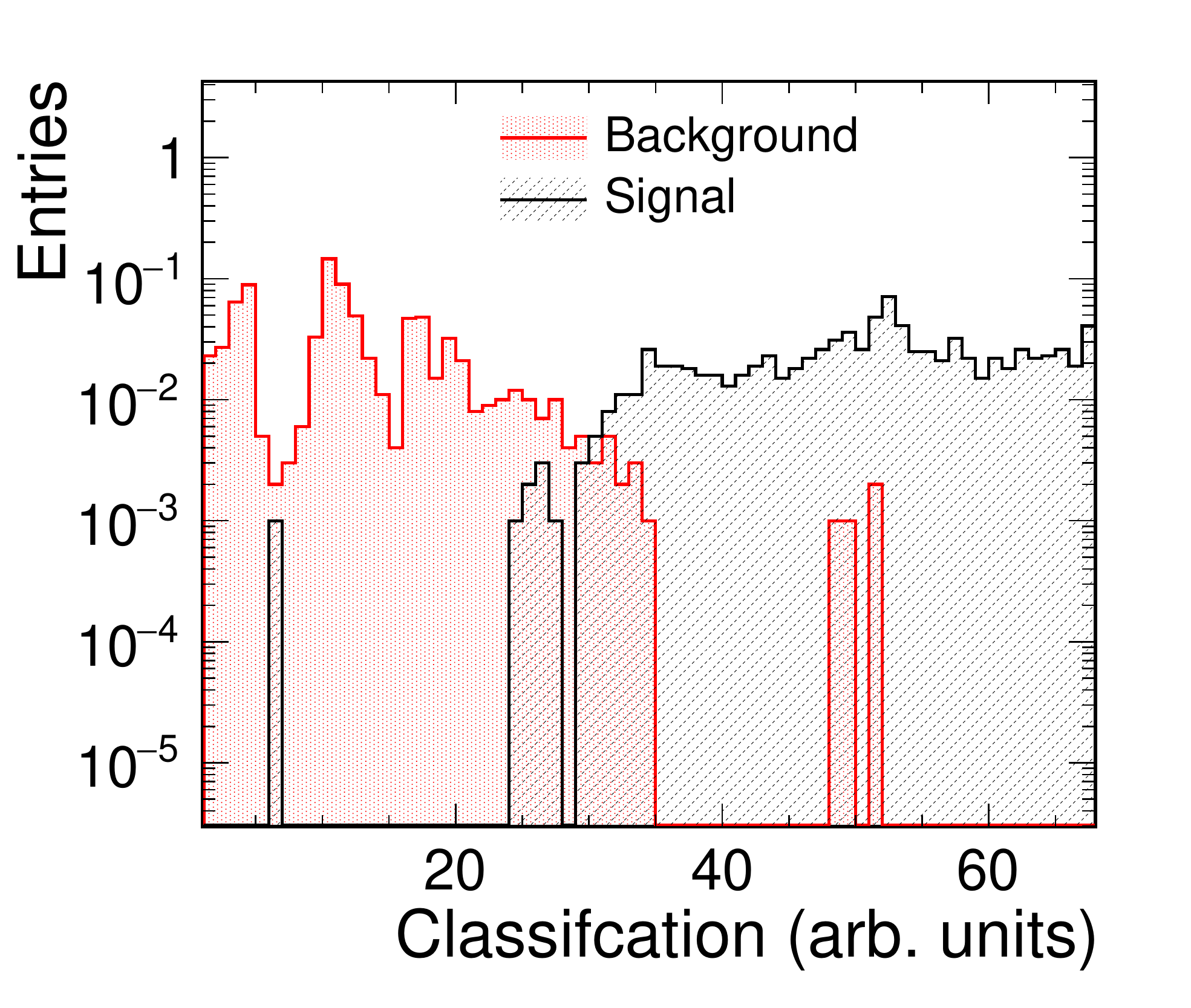}%
	\caption{SOM classification distribution for signal and background.}
	\label{fig:kohonen_distanceAndtrainedMap}
\end{figure}

The trained one-dimensional grid serves as classification axis and is displayed in  Fig.~\ref{fig:kohonen_distanceAndtrainedMap} for the evaluation set.
The ROC curve exhibits a similar performance as the MLP, as illustrated in Fig.~\ref{fig:rocOverlay_202109}.
While the AUC is $98.15^{+0.04}_{-0.04}$\,\% and therefore lower than for the MLP, the signal efficiency at  $\epsilon_S(\epsilon_B = 10^{-4})$ background rejection evaluates to $56.8^{+7.1}_{-8.8}$\,\% , which is superior to the MLP.

\subsection{Autoencoders (AE)}

\begin{figure}[tbp]
	\centering
	\includegraphics[width=\columnwidth]{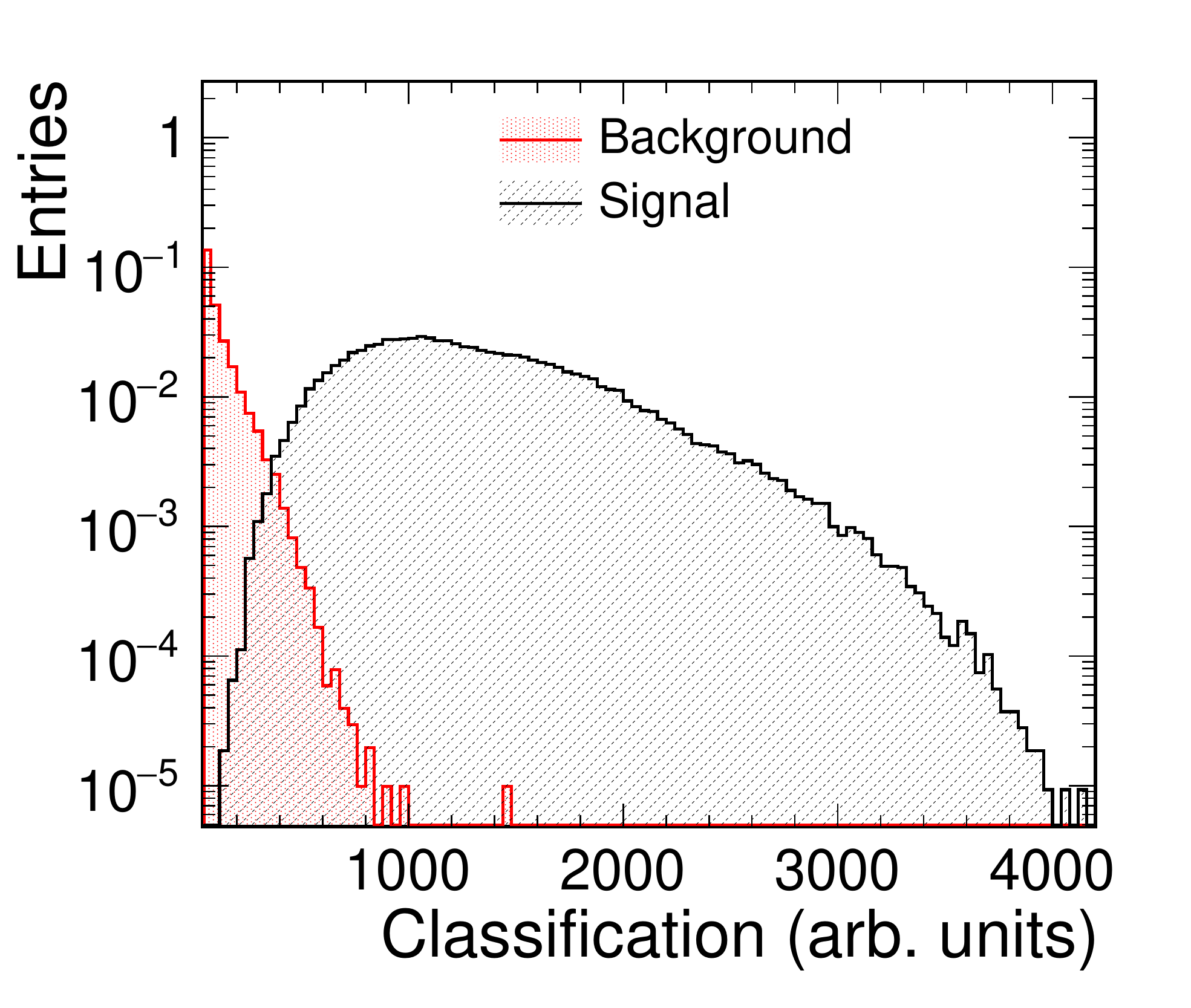}%
	\caption{AE classification distribution for signal and background.}
	\label{fig:autoencoder_classLoss}
\end{figure}

\begin{table*}[t]
	\centering
	\begin{tabular}{ll}
		\toprule
		\textbf{Parameter} & \textbf{Value}  \\ 
		\midrule
		Activation function & Rectified Linear Unit (ReLu) \\
		Loss function & Mean squared error  \\
		Optimizer & Adaptive Moment Estimation (Adam) \\
		Learning rate & 1e-4 \\
		\bottomrule
	\end{tabular}
	\caption{Hyperparameters used for the Autoencoder.}
	\label{tab:parametersAE}
\end{table*}

An Autoencoder (AE) is a feed-forward multilayer neural network that aims to reproduce the input vector without using an identity mapping. 
It consists of two parts: an encoder and decoder.
While the encoder compresses the input to a lower-dimensional vector, the decoder reconstructs the original input from the reduced representation. 
The latent space in the centre of the AE is an information bottleneck that enforces the selection of relevant patterns from the input data. 

During training, only background events are presented to the AE and their reconstruction error is minimised, making the AE specialized in the reproduction of background events.
In the evaluation phase, the AE is able to recognize background events by a low reconstruction error. 
Signal events appear anomalous to the AE and are characterised by a high error, that can therefore serve as an anomaly score~\cite{an2015variational}.

The same number of background events for training and testing as for the MLP are used. 
The hyperparameters of the AE are listed in Tab~\ref{tab:parametersAE} and the architecture is shown in the Appendix.
There, the robustness of the AE against e.g. impurities in the training set is discussed as well.

The classification distribution for the evaluation set is displayed in Fig.~\ref{fig:autoencoder_classLoss}.
While there is a considerable overlap region for low reconstruction errors, a high signal purity is observable at high values.
The resulting ROC curve of the AE is presented in Fig.~\ref{fig:rocOverlay_202109}.
The AUC evaluates to $98.86^{+0.05}_{-0.02}$\,\%  and is therefore worse compared to the MLP but superior to the SOM.
The signal efficiency at high background rejection levels still reaches $60.1^{+3.3}_{-2.7}$\,\%  showing that the AE outperforms the other two algorithms.

\section{Summary and Outlook}

The unsupervised identification of anomalous pixel detector data using Self-Organizing Maps and Autoencoders was presented. 
To exemplify the approach, hypothetical magnetic monopoles at the Belle II pxdetector were simulated and identified against background data.
The two unsupervised algorithms have shown similar performance as a convolutional Multilayer Perceptron.
Both of them outperform the Multilayer Perceptron, if high background rejection levels are required, with the Autoencoder being a few percent better than the Self-Organizing Maps. 

This study is an essential cornerstone for future online applications of anomaly detection at the Belle\,II pixel detector in order to further improve its sensitivity to undiscovered physics.

\section{Acknowledgements}

The authors would like to thank the Belle II Collaboration and Belle II PXD group for their support.
We are grateful to Klemens Lautenbach, Carsten Niebuhr and Maiko Takahashi for their valuable input on the simulation framework and for useful discussions and suggestions.  
This work was supported by the Bundesministerium für Bildung und Forschung within the joint research project 05H2021 (ErUM-FSP T09) under grant agreements 05H19RGKBA and 05H21RGKB1.

\section{Data Availability Statement}

The data that support the findings of this study are available from the Belle\,II collaboration but restrictions apply to the availability of these data, which were used under licence for the current study, and so are not publicly available. 
Data are however available from the authors upon reasonable request and with permission of the Belle\,II collaboration.

\bibliographystyle{sn-mathphys}    
\bibliography{sn-bibliography.bib}

\appendix

\section{Neural Network Architecture}

\begin{table*}[h]
	\centering
	\begin{tabular}{lllll}
		\toprule
		\textbf{Layer} & \textbf{Kernel} & \textbf{Padding} & \textbf{Nodes} & \textbf{Activation} \\
		\midrule
		Convolutional 2D & $3 \times 3$ & $1 \times 1$ & 84 & Leaky ReLU \\
		Convolutional 2D & $3 \times 3$ & $1 \times 1$ & 128 & Leaky ReLU \\
		Convolutional 2D & $3 \times 3$ & $1 \times 1$ & 256 & Leaky ReLU \\
		Linear & - &  - & 128 & Leaky ReLU \\
		Linear & - &  - & 32 & Leaky ReLU \\
		Linear & - &  - & 1 & - \\
		\bottomrule
	\end{tabular}
	\caption{Network architecture of the MLP.}
	\label{tab:mlp_architecture}
\end{table*}

\begin{table*}[h]
	\centering
	\begin{tabular}{lll}
		\toprule
		\textbf{Layer} & \textbf{Nodes} & \textbf{Activation} \\
		\midrule
		Linear & 84 &  ReLU \\
		Linear & 64 &  ReLU \\
		Linear & 16 &  ReLU \\
		Linear & 8 &  ReLU \\
		Linear & 16 &  ReLU \\
		Linear & 64 &  ReLU \\
		Linear & 84 &  ReLU \\
		\bottomrule
	\end{tabular}
	\caption{Network architecture of the AE.}
	\label{tab:ae_architecture}
\end{table*}
%

\section{Robustness}

\subsection{Signal in training data}

We consider the hypothetical scenario of having a background training set that is contaminated with signal events. 
In this case, the AE is already exposed to signal during training and will consequently adapt to it, potentially resulting in a lower reconstruction error for signal and therefore a weaker classification performance. 
To evaluate the impact of a contaminated data set, 0.01\,\% / 0.1\,\% / 1\,\% of the training events are replaced by signal and the analysis is repeated.

\begin{figure}[tbp]
	\centering
	\includegraphics[width=\columnwidth]{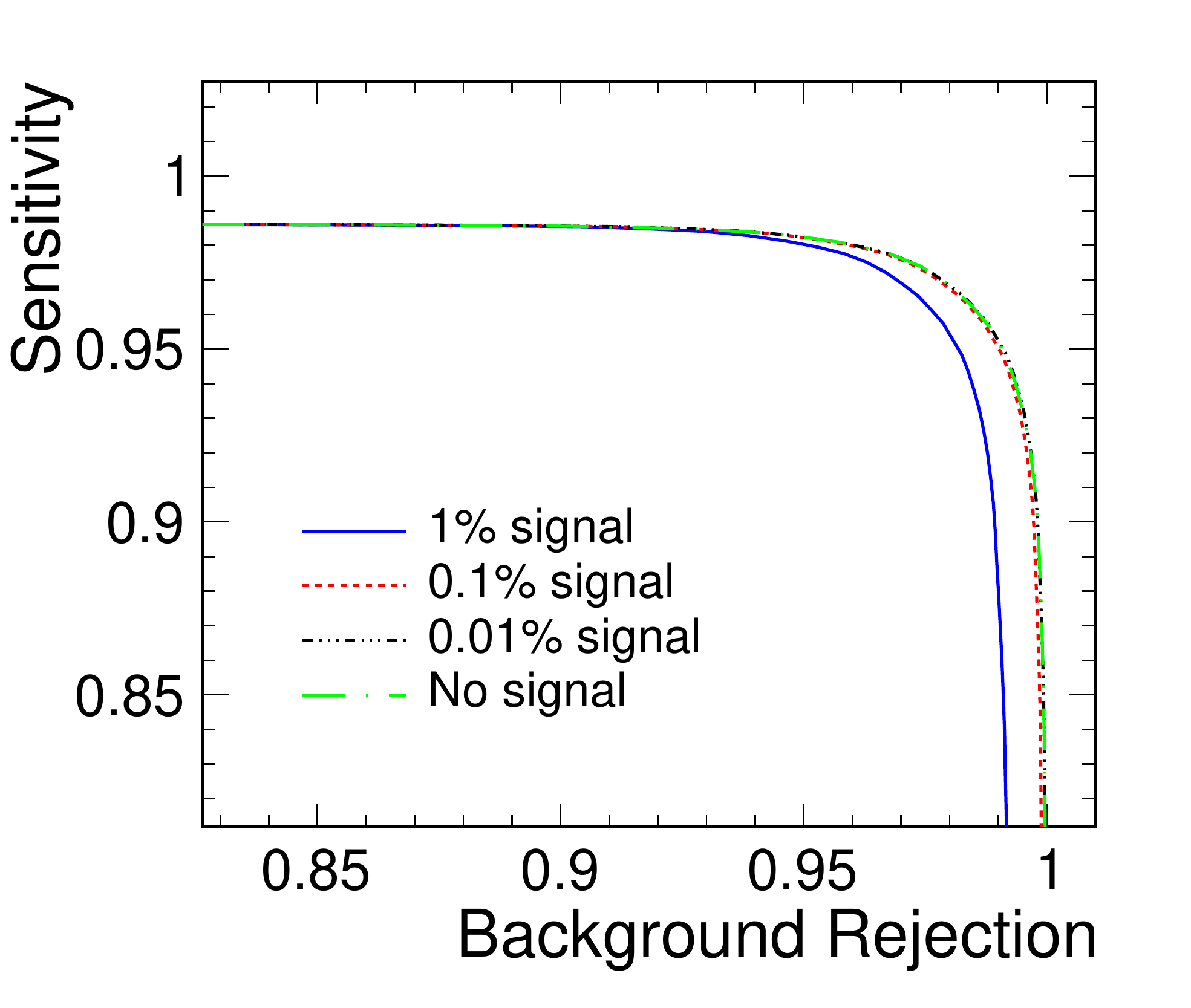}%
	\caption{ROC curves for different signal levels in the training sample.}
	\label{fig:roc_diffSignalnBgd}
\end{figure}

The resulting ROC curves are presented in Fig.~\ref{fig:roc_diffSignalnBgd}.
The performance of the AE deteriorates at 1\,\% signal in the data but is not affected for the lower contamination values. 
It can be concluded that a signal component in the training set has to be sufficiently rare to guarantee the optimal performance of the AE.

\subsection{Size of Pixel Matrix}
\label{sec:sizePixelMatrix}

\begin{figure}[tbp]
	\centering
	\includegraphics[width=\columnwidth]{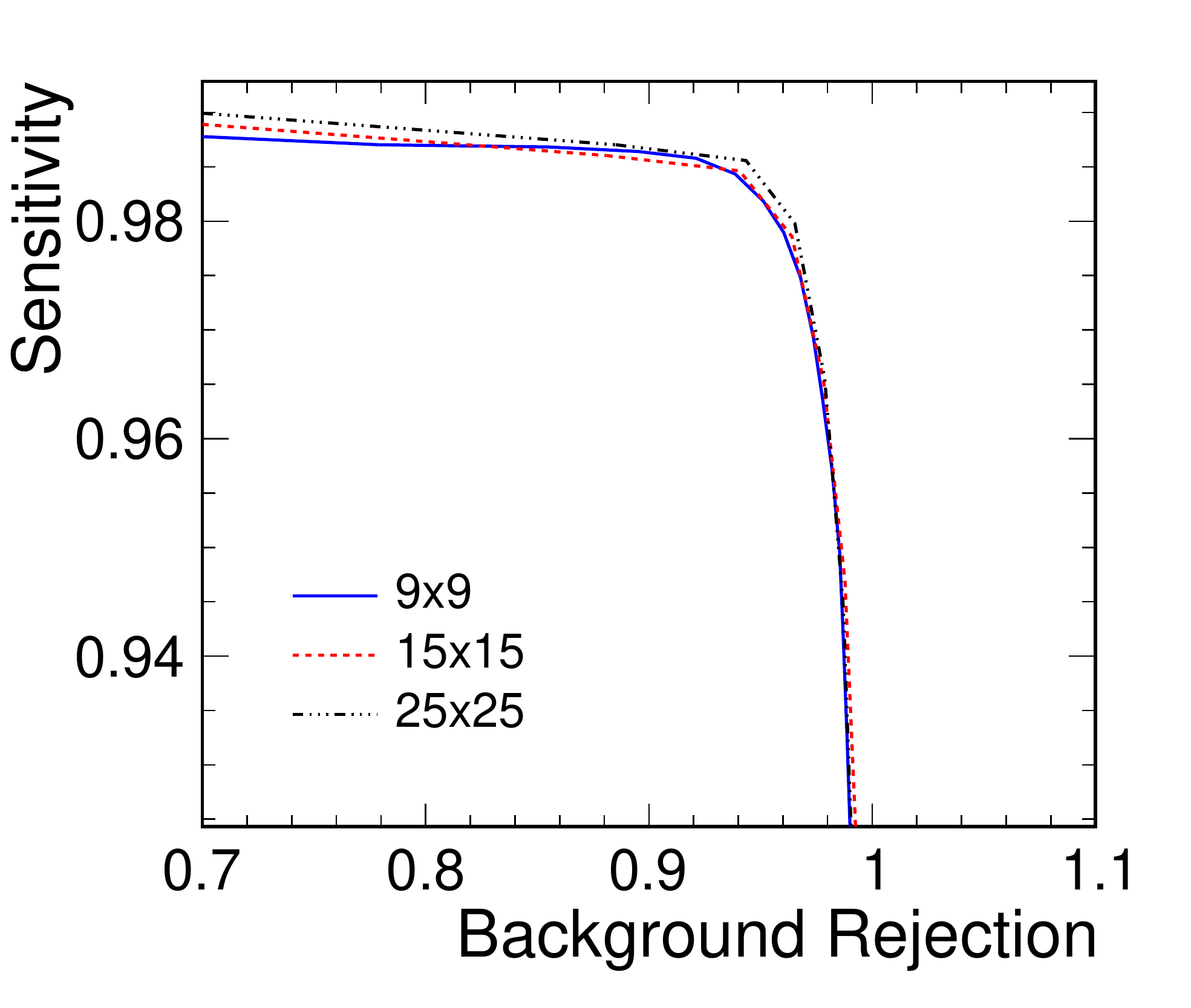}%
	\caption{ROC curves using different matrix sizes for the input data.}
	\label{fig:roc_diffPxlMatrix}
\end{figure}

To investigate the impact of the size of the pixel matrix, larger matrices are extracted from the same data and simulation sets. 
The training, testing and evaluation steps of the AE are repeated and three ROC curves belonging to matrix sizes of 9\,$\times$\,9,  15\,$\times$\,15 and 25\,$\times$\,25 are computed, as illustrated in Fig.~\ref{fig:roc_diffPxlMatrix}.
An improvement for larger matrix sizes at low background rejection levels is observable. 
As the improvement is comparably small, the increase in computation time is not justified for this study and therefore the 9\,$\times$\,9 matrix is kept. 

\subsection{Simulated Background}

\begin{figure}[tbp]
	\centering
	\includegraphics[width=\columnwidth]{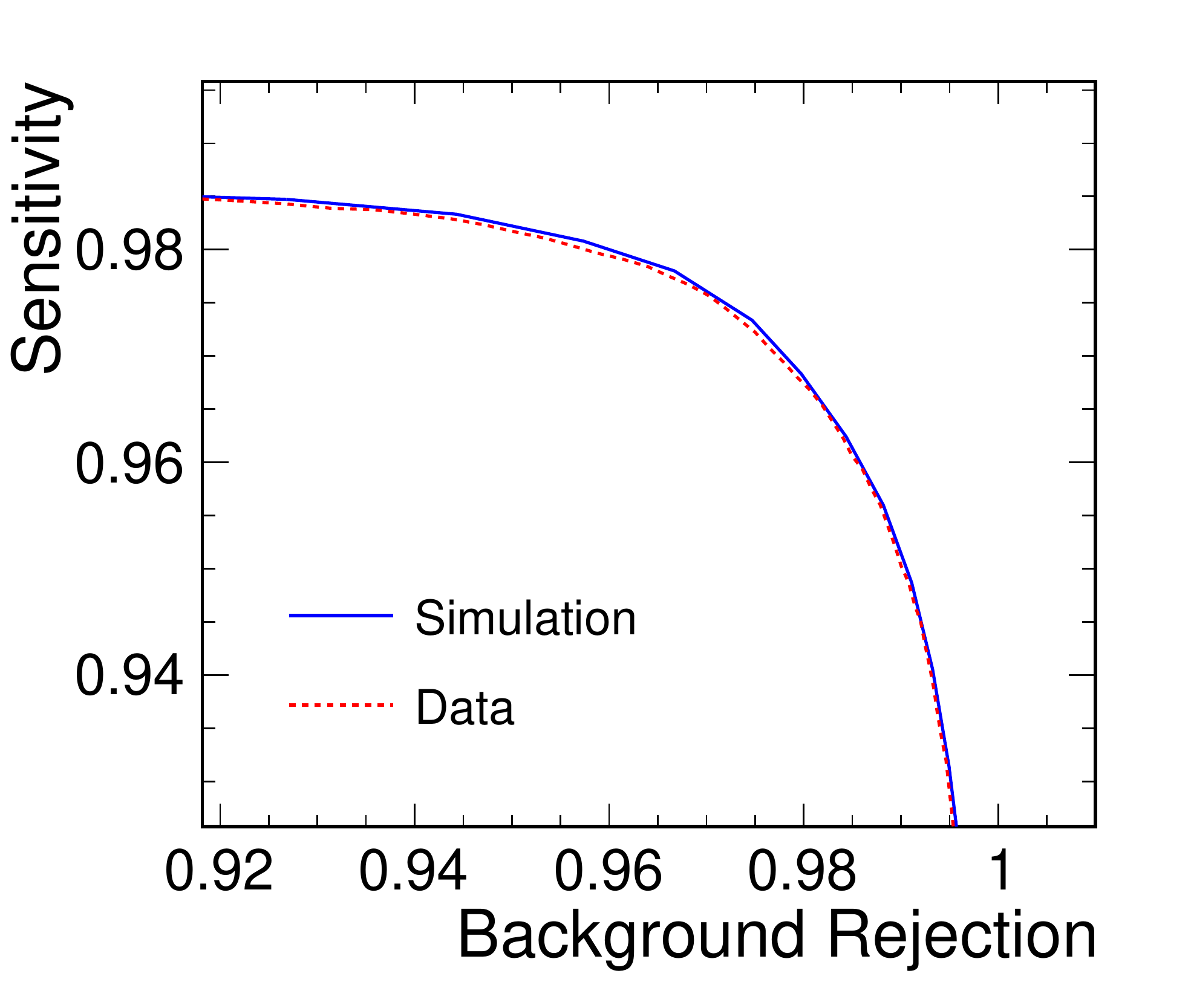}%
	\caption{ROC curves for simulated and real beam background.
		In both cases, the same signal set is used.}
	\label{fig:roc_dataSim}
\end{figure}

Data from the PXD is subject to various systematic effects that were not explicitly accounted for in this study such as particularly noisy pixels or pixel-by-pixel variations resulting in a slightly different AUC value for the same energy deposition. 
In online applications, correcting for these effects is not easily feasible.
It is therefore beneficial to examine their impact on the classification performance. 
To this end, dedicated beam-background simulations are performed.
The extraction of pixel matrices is conducted in the same manner as for data and an AE is trained and evaluated on the simulation. 

The resulting ROC curve is compared to the one from data in Fig.~\ref{fig:roc_dataSim}.
Good agreement between the two curves is achieved, which implies an excellent background modelling in simulation.
In addition, remaining differences between data and simulation related to noisy pixels or pixel-by-pixel variations do not affect the results.
Although noisy pixels have the potential to disturb the training, the results from the previous section imply that a disturbance of about \SI{1}{\percent} of the considered matrices is necessary to affect the AE.
While the pixel-by-pixel variations introduce uncertainties in the ADC values, the large difference in energy deposition between background and signal are not sensitive to these comparably small variations.

\end{document}